\documentclass[%
reprint,
 amsmath,amssymb,
 aps,
 pre,
floatfix
]{revtex4-1}

\usepackage{graphicx}
\usepackage{dcolumn}
\usepackage{bm}
\usepackage{url}
\usepackage{hyperref}
\usepackage{pslatex} 
\usepackage[ruled,longend]{algorithm2e}
\usepackage{color}

\begin{document}

\preprint{APS/123-QED}
\title{Infostop: Scalable stop-location detection in multi-user mobility data.}

\author{Ulf Aslak}
    \email{ulfaslak@gmail.com}
    \affiliation{
        Centre for Social Data Science, University of Copenhagen, DK-1353 K\o{}benhavn K
    }
    \affiliation{
        DTU Compute, Technical University of Denmark, DK-2800 Kgs.~Lyngby
    }
\author{Laura Alessandretti}
    \affiliation{
        Centre for Social Data Science, University of Copenhagen, DK-1353 K\o{}benhavn K
    }
    \affiliation{
        DTU Compute, Technical University of Denmark, DK-2800 Kgs.~Lyngby
    }

\date{\today}

\begin{abstract}
Data-driven research in mobility has prospered in recent years, providing solutions to real-world challenges including forecasting epidemics and planning transportation. These advancements were facilitated by computational tools enabling the analysis of large-scale data-sets of digital traces. One of the challenges when pre-processing spatial trajectories is the so-called \emph{stop location detection}, that entails the reduction of raw time series to sequences of destinations where an individual was stationary. The most widely adopted solution was proposed by Hariharan and Toyama (2004) and involves filtering out non-stationary measurements, then applying agglomerative clustering on the stationary points. The state-of-the-art method, however, suffers of two limitations: (i) frequently visited places located very close (such as adjacent buildings) are likely to be merged into a unique location, due to inherent measurement noise, (ii) traces for multiple users can not be analysed simultaneously, thus the definition of destination is not shared across users. In this paper, we describe the \textit{Infostop} algorithm for stop location detection that overcomes the limitations of the state-of-the-art solution by leveraging the flow-based network community detection algorithm Infomap. We test Infostop for a population of $\sim 1000$ individuals with highly overlapping mobility. We show that the size of locations detected by Infostop saturates for increasing number of users and that time complexity grows slower than for previous solutions. We demonstrate that Infostop can be used to easily infer social meetings. Finally, we provide an open-source implementation of Infostop, written in Python and C++, that has a simple API and can be used both for labeling time-ordered coordinate sequences (GPS or otherwise), and unordered sets of spatial points.
\end{abstract}

\maketitle

\section{Introduction}

Understanding human mobility is paramount to tackle real-world challenges, from containing epidemics to modelling transportation \cite{barbosa2018human}. Since the end of the $19th$ century, studies on individual and collective human movements were carried out in a broad range of disciplines\cite{barbosa2018human}. Until recently, these analyses relied on data collected from census, surveys and self-reported travel diaries, which suffered from limitations including limited resolution, small sample sizes and reporting biases \cite{barbosa2018human}. In the last decade, the introduction and diffusion of mobile phone devices and other positioning technologies have renewed the interest of the scientific community for human mobility. Today, trajectories can be inferred from sources including mobile phone call logs, location based social networks information, and data collected from GPS devices \cite{barbosa2018human}. Using data collected by electronics, media and telecommunication companies \cite{blondel2012data} or by running large-scale experiments\cite{stopczynski2014measuring}, scientists can base their studies on large datasets collecting the positions of millions of individuals over months and years, at the resolution of meters and seconds. These novel datasets offer incredible new possibilities, but also call for novel tools.
    
One of the key challenges related to the pre-processing of digital traces is translating raw sequences of spatial coordinates into comprehensible data. Typically, raw data is a stream of records (\textit{id}, \textit{lat}, \textit{lon},\textit{t}), where \textit{id} is the identifier of a phone or other positioning object, and \textit{lat} and \textit{lon} characterize its position in space at time \textit{t}. Due to inherent measurement noise, \textit{lat} and \textit{lon} can fluctuate even when the sensor is perfectly still. In this format, the data is far from being simply understood. In fact, we think of human trajectories as sequences of \emph{trips} between destinations and \emph{stays} at destinations. For example, Alice was at home today from $12 am$ to $7:30 am$, she then went to the office, where she stayed from $8:00am$ to $5pm$, and then went back home, where she has been from $5:30 pm$ until $11:59 pm$. Understanding individual trajectory as sequences of \emph{stays} and \emph{trips} is the first step necessary to compute relevant metrics that characterize mobility behaviour, including the total distance an individual has travelled, the radius of gyration, or the number of unique locations visited \cite{barbosa2018human}.

Several solutions were proposed to transform streams of spatio-temporal records into \emph{stays} and \emph{trips} sequences~\cite{Gong2015,Cao2010,montoliu2010discovering,do2013places,wan2013life,zhao2015stop}. To date, the most adopted solution is the `Lachesis Project' developed by Hariharan and Toyama (2004) \cite{hariharan2004project}. The algorithm is designed to process one individual trace at the time and it is based on a two-steps procedure. (i) First, it identifies \emph{stays} as periods when an individual does not stray further than a maximum distance $r_1$ for a minimum duration $t_{min}$, where $r_1$ and $t_{min}$ are parameters. Each stay is described as a tuple ($start$, $end$, $lat$, $lon$), such that an individual was stationary between $start$ and $end$ at the location identified by latitude and longitude $(lat, lon)$. (ii) Then, it uses spatial clustering, with parameter $r_2$, to cluster the coordinates $(lat, lon)$ of all stays, in order to identify \emph{destinations}. This step attaches a label to each of the \emph{stays} identified in the previous step. For example, if Alice has two \emph{stays}, one in the morning and one in the evening, that are located closer than $r_2$ from each other, they will be identified as two stays in the same destination (see \cite{hariharan2004project} for more details). In the most widely used formulation of the algorithm, scientists use the DBSCAN clustering algorithm \cite{ester1996density} to perform step (ii) \cite{pappalardo2019scikit}. 
    
The state-of-the-art solution based on DBSCAN suffers of two major limitations. First, if an individual frequently stops at two separate locations that are near each other (e.g. adjacent buildings on a campus), location measurements from all visits should fall into two separate clusters, but due to sampling noise they may overlap. In fact, clustering algorithms that operate in euclidean space, such as DBSCAN~\cite{ester1996density} or other agglomerative clustering methods are not well suited for this problem, as clusters that overlap tend to get merged. Second, although it would be possible, in principle, to use DBSCAN to cluster stays of multiple users, DBSCAN is not suitable for multi-user traces, especially when individuals share many destinations. This is again due to the fact that DBSCAN aggregates areas dense in locations into single large locations.
    
In order to overcome the aforementioned limitations, we introduce the Infostop algorithm for stop-location detection, described in section \ref{sec:algorithm}. Infostop is a two-step algorithm: the first step corresponds to step (i) of the Hariharan and Toyama algorithm, the second step leverages the Infomap clustering algorithm for networks to allow for the the detection of overlapping clusters and provide fast and scalable multi-user stop-location detection.
    
\begin{figure}
    \centering
    \includegraphics[width=\linewidth]{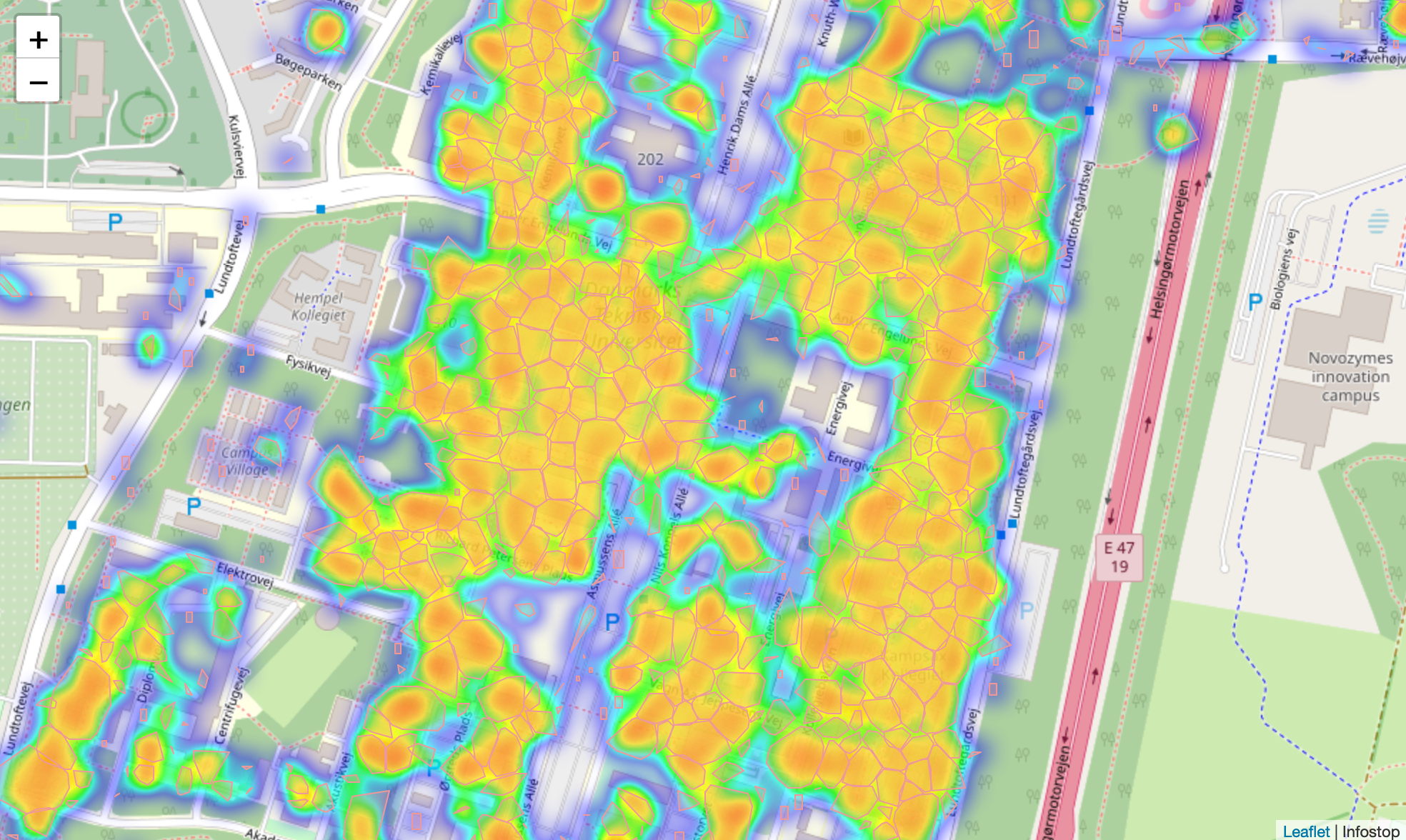}
    \caption{\textbf{Infostop identifies compact locations in dense areas.} Heatmap showing the density of points at DTU University Campus. Polygons represent the \emph{destinations} identified by Infostop. }
    \label{fig:heatmap}
\end{figure}

\begin{figure}
    \centering
    \includegraphics[width=\linewidth]{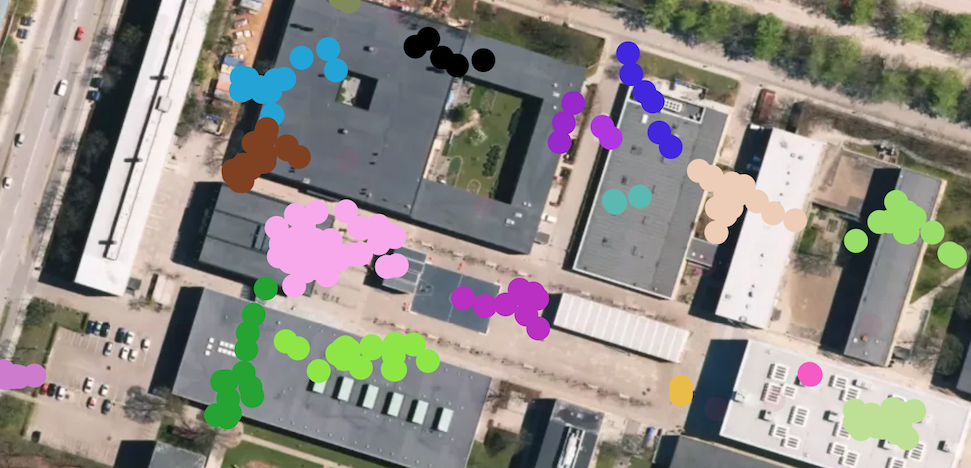}
    \caption{\textbf{Example of locations detected with Infostop.} A sample of labeled locations from an anonymous user on a university campus. Only the median of each stay is shown. It is apparent that many of the clusters are either very near or even overlap to some degree.}
    \label{fig:infostop_example_map}
\end{figure}

\begin{figure}
    \centering
    \includegraphics[width=\linewidth]{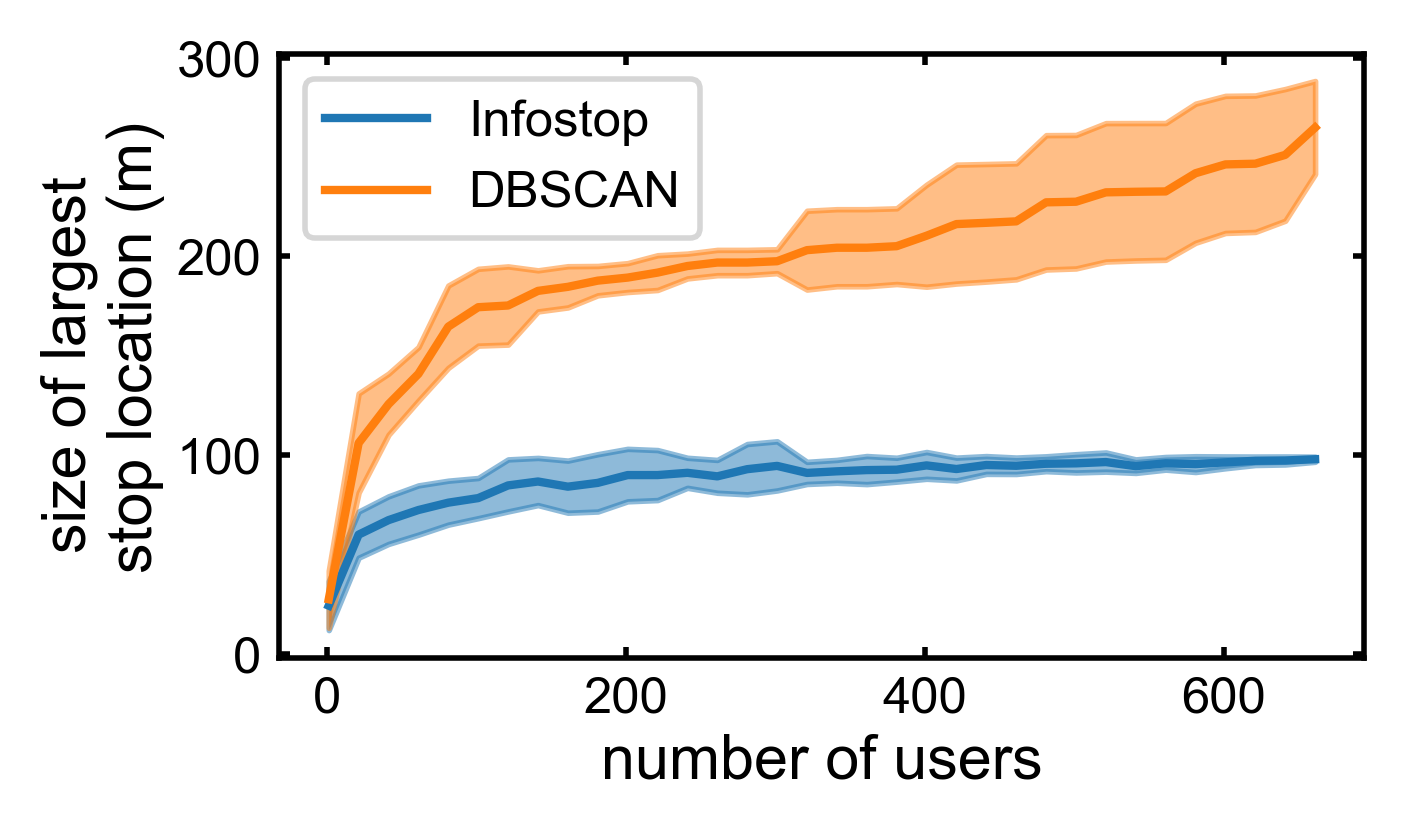}
    \caption{\textbf{Destinations identified by infostop are compact.} The size of the largest destination identified by Infostop (blue line) and the DBSCAN state-of-the-art algorithm (orange line) for increasing number of users in the dataset. Results are shown for parameters $r1=20 m$, $r2=20 m$, $t_{min}=10min$, $t_{max}=24 h$, and using one month of data (22555616 records). Errorbars are computed across 30 different instances (by selecting different sets of users).}
    \label{fig:space}
\end{figure}

In section \ref{sec:results}, we show that Infostop allows the identification of compact locations for large dataset of individuals with largely overlapping destinations (see Fig.~\ref{fig:heatmap} and \ref{fig:infostop_example_map}), and, in this respect, it performs better than state-of-the-art methods (see Fig.~\ref{fig:space}). We find that multi-user solutions are comparable to individual-user solutions (see Fig.~\ref{fig:rand_score}). We demonstrate that Infostop has lower time complexity than the state-of-the-art (see Fig.~\ref{fig:time}). We show that Infostop can be used to identify social contacts. In section \ref{sec:implementation}, we present a fast and flexible open-source implementation of Infostop available in Python.

\section{The Infostop Algorithm  \label{sec:algorithm}}

Infostop detects stop-locations from raw sequences of spatio-temporal data. Like the method in \cite{hariharan2004project}, Infostop works in two-steps. First, it identifies \emph{stays}, and then \emph{destinations}.

The algorithm takes the following input.\\

\textbf{Data.} A collection of sequences of records. Each sequence $T_u$ describes the movement of a different individual $u$, and consists of $N_u$ records $[ (t_{u,i}, x_{u,i}, y_{u,i})]$, where $t_{u,i}$ is time, $x_{u,i}$ and $y_{u,i}$ are coordinates in a two-dimensional space. Note that sequences are ordered by time and each of them can have a different length.\\
    
\textbf{Parameters.}
\begin{itemize}
    \item $r_1$, the maximum roaming distance allowed for two points within the same stay.
    \item $t_{min}$, the minimum duration of a stay.
    \item $t_{max}$, the maximum time difference between two consecutive records for them to be considered within the same stay.
    \item $r_2$, the typical distance between two stays in the same destination.
\end{itemize}

The algorithm works in two steps. In step (i), Infostop identifies \emph{stays} and \emph{trips}, for each of the trajectories in the collection $T_u$, and assigns to each record a label in \{trip, stay\}. Here, Infostop follows closely the method developed by \cite{hariharan2004project}. For more details on the implementation of step (i), the reader can refer to the code in~\footnote{\url{https://github.com/ulfaslak/infostop/blob/master/cpputils/main.cpp}}.

In step (ii), points are clustered using Infomap. First, we build a network, where nodes correspond to \emph{stays} (as identified by step (i)), and a link exists between two nodes if they are located at distance smaller than $r_2$. Note that we include in the network \emph{stays} deriving from all the individual trajectories. Then, we find \textit{communities} of nodes using the Infomap algorithm~\cite{rosvall2009map}. These communities identify \emph{destinations}, in a definition that is shared across individuals. For more details on the implementation of step (ii), the reader can refer to the code in\footnote{\url{https://github.com/ulfaslak/infostop/blob/master/infostop/utils.py}}. 

Thus, for each sequence of records $T_u$ in the input, Infostop returns a sequence of labels with corresponding length. Records identified as trips are assigned label $l_{u,i}=-1$, while records that are identified as stays are assigned a positive integer, identifying the corresponding \emph{destination}. 

\section{Results \label{sec:results}}
We test the algorithm on the trajectories of $\sim 1000$ students at the Technical University of Denmark that were part of the same study program and thus have largely overlapping mobility trajectories \cite{stopczynski2014measuring}. Data was collected as part of an experiment that took place between September $2013$ and September $2015$. Participants’ position over time was estimated from a combination GPS and WiFi information, resulting in samples every $1-2$ minutes. The location estimation error was below $50$~m in $95\%$ of the cases. 

\begin{figure}
    \centering
    \includegraphics[width=\linewidth]{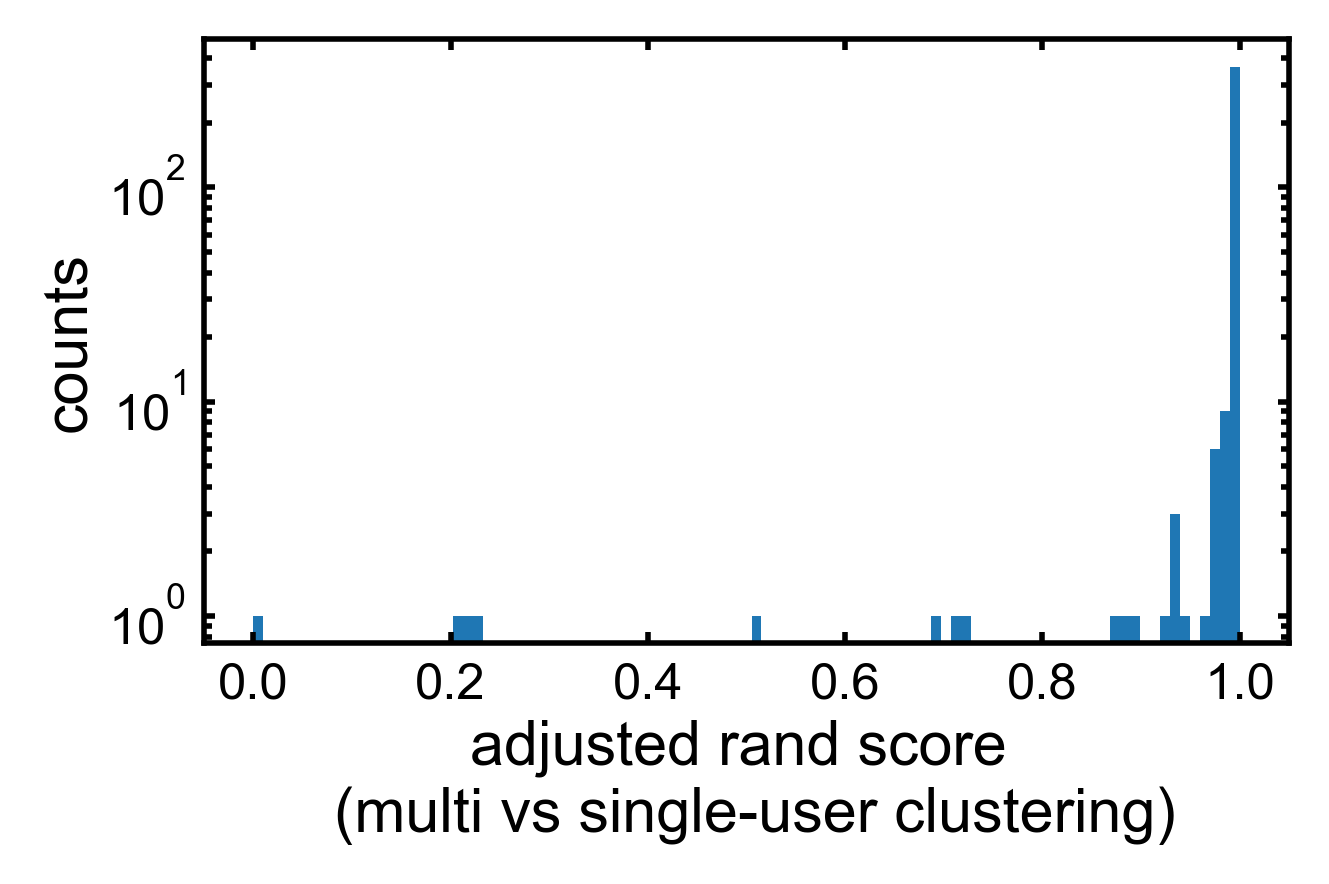}
    \caption{\textbf{Similarity of the single-user vs multi-user solution.} The distribution of the adjusted rand score across individuals for the single-user vs the multi-user solution of Infostop. The rand score assesses the similarity of two partitions (it is equal to one for identical solutions and to zero in the random case).  Results are shown for parameters $r1=20 m$, $r2=20 m$, $t_{min}=10min$, $t_{max}=24 h$.}
    \label{fig:rand_score}
\end{figure}

\begin{figure}
    \centering
    \includegraphics[width=\linewidth]{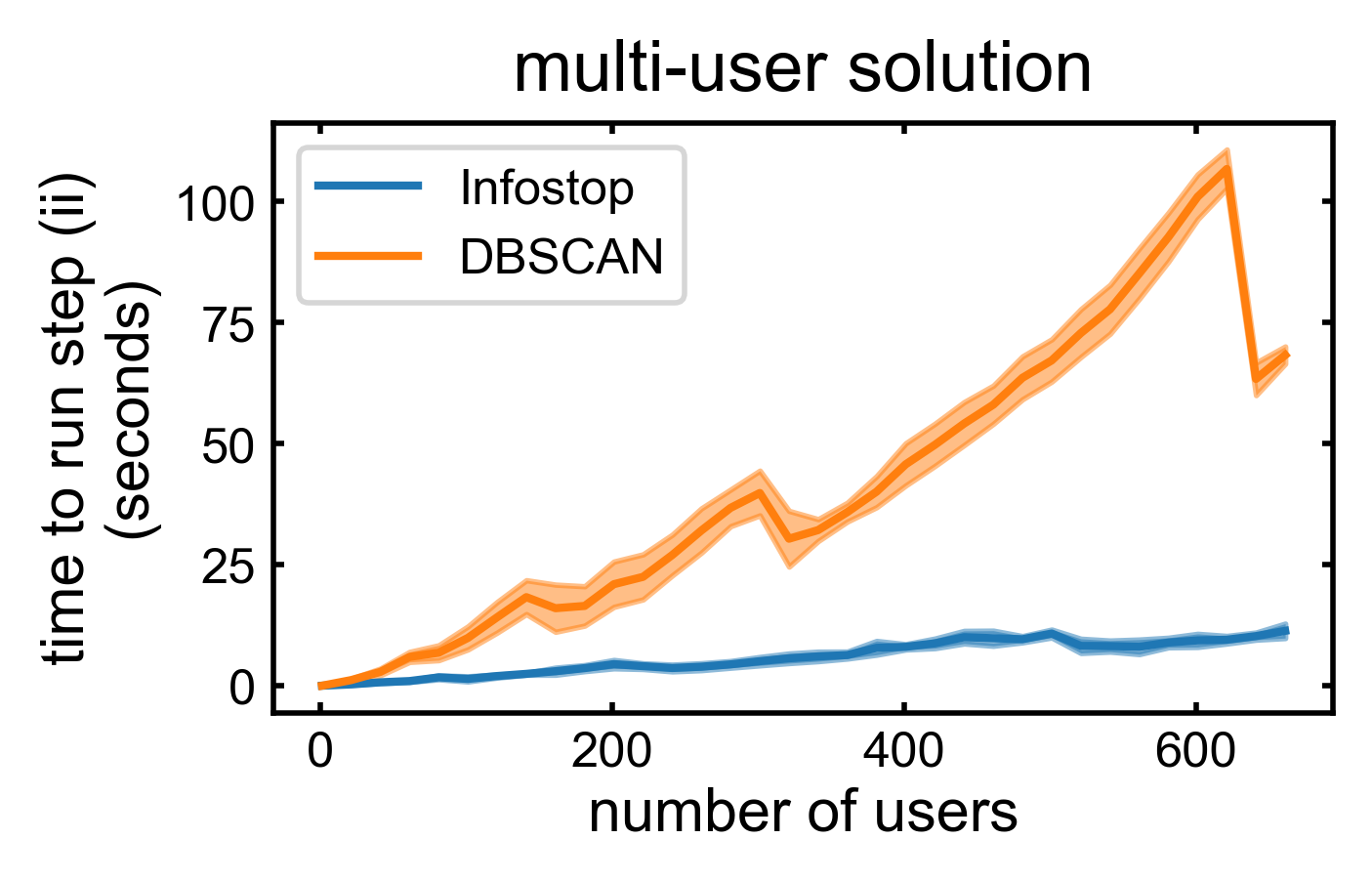}
    \includegraphics[width=\linewidth]{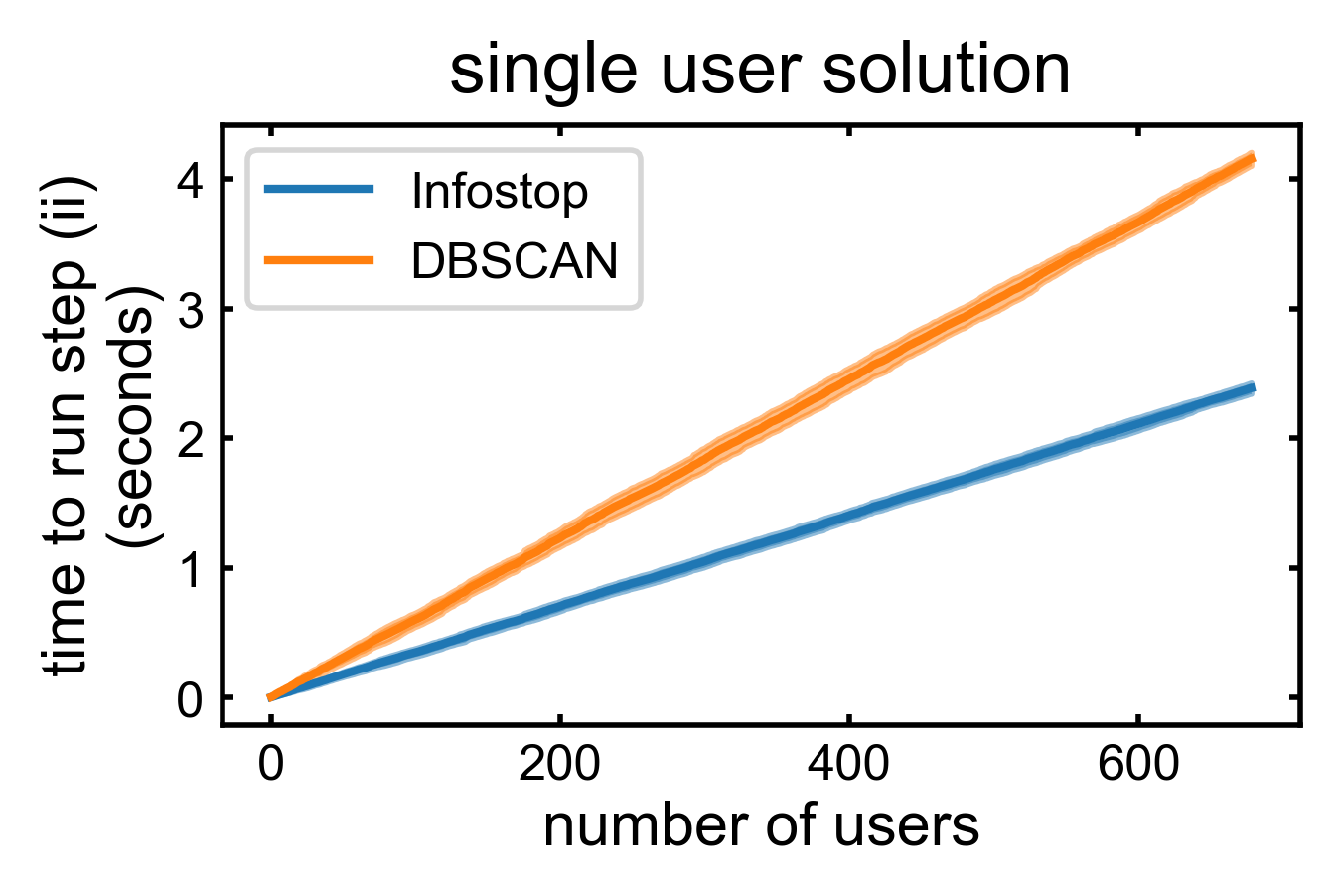}
    \caption{\textbf{Infostop is faster than the state-of-the-art solution.} The time in seconds to run Infostop (blue line) and DBSCAN (orange line) on the same machine for increasing number of points in the dataset. Results are shown for parameters $r1=20 m$, $r2=20 m$, $t_{min}=10min$, $t_{max}=24 h$, and using one month of data (22555616 records). Results are shown for the multi-user solution (top) and the single-user solution (bottom).}
    \label{fig:time}
\end{figure}

First, we test the accuracy of the multi-user version of Infostop, by comparing the labels obtained by runnning stop-detection for each user separately vs for all users simultaneously. We find that the median value of the adjusted rand index~\cite{steinley2004properties} between the two solutions is equal to one, revealing that the single-user and multi-user solutions are almost identical (see Fig.~\ref{fig:rand_score}). 
Then, we assess the quality of the Infostop solution, by measuring the size of the largest destination (maximum distance between any two points with the same destination label) as a function of the number of users included in the dataset. We find that, for a choice of $r_2=20 m$, the quantity saturate at $\sim100 m$, while the solution of DBSCAN, run with the same distance parameter, diverges. Then, we measure the time it takes to run step (ii) of the stop-location detection algorithm using Infomap or DBSCAN on the same machine. We use the Python wrapper for Infomap by Daniel Edler, Anton Eriksson and Martin Rosvall \footnote{\url{https://mapequation.github.io/infomap/python/}}, and the the scikit-learn Python implementation \cite{scikit-learn} of DBSCAN with parameters \texttt{n\_jobs=1}, \texttt{metric=haversine}, \texttt{leaf\_size=40}, \texttt{algorithm=ball\_tree}. We find that Infomap runs faster compared to DBSCAN (see Fig.~\ref{fig:time}).

\begin{figure}
    \centering
    \includegraphics[width=\linewidth]{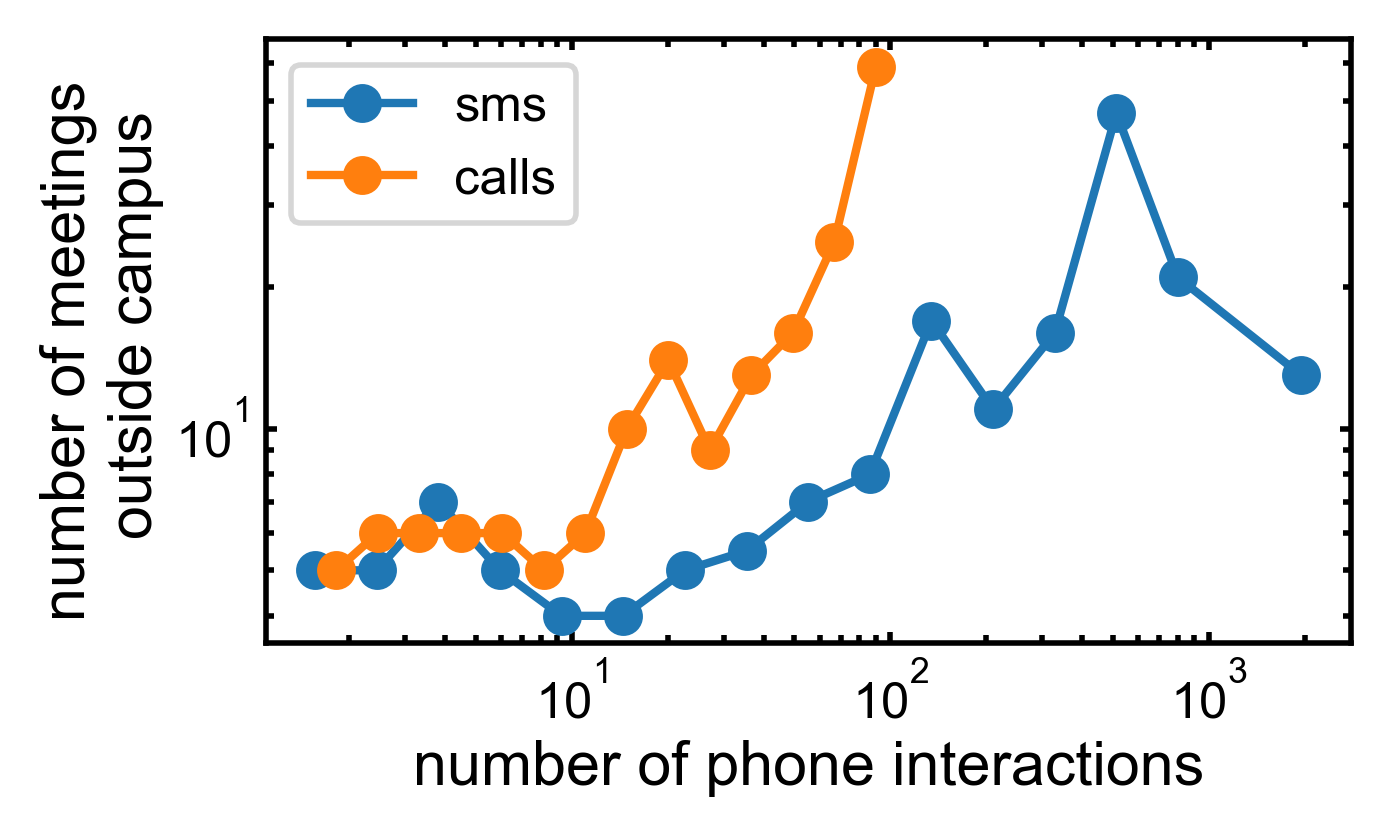}
    \caption{\textbf{Infostop allows to identify social meetings.} The median number of meetings outside campus, or situations where users spent more than $10$ minutes simultaneously in the same destination vs the number of mobile phone interactions, by sms (orange line) and calls (blue line). Results are aggregated over logarithmic bins. }
    \label{fig:interactions}
\end{figure}

Finally, we use the outcome of the algorithm to identify social contacts between individuals. Since Infostop identifies shared destinations across users, it makes it fast to identify situations when users spent simultaneously time in the same destination. We find that the number of such instances, after removing meetings occurring in the DTU campus, correlates positively with the number of calls and sms exchanged by two individuals (see Fig.~\ref{fig:interactions}). 
    
\section{Open Source Implementation  \label{sec:implementation}}
We developed an implementation of Infostop in C++ and Python. The code is open-sorce, it is wrapped in Python, and can be installed using \texttt{pip}. The package is located at\\ \url{https://github.com/ulfaslak/infostop}.

The \texttt{Infostop} package implements two models as separate classes.

\begin{enumerate}
    \item \texttt{Infostop}: \textbf{Infer stop-location labels from mobility traces.} An instance of the class is initialised using the parameters $r_1$, $t_{min}$, $t_{max}$, $r_2$ described in Section \ref{sec:algorithm}. The main method of the class is \texttt{fit\_predict}, which takes as input a collection of trajectories and returns the label associated to each record. The method works in multiple steps: (a) Temporal downsampling: detect stationary points and store the median coordinates of each of them. (b) Spatial downsampling: remove duplicate coordinates, optionally with prior downsampling to total lower number of points to be clustered. (c) Using a ball tree algorithm, identify stationary events that are closer than $r_2$ and (d) constructs a network of stationary events (where a link exists between two nodes if the are closer than ${r_2}$), and cluster the network with the Infomap algorithm. (e) Reverse spatial downsampling. (f) Reverse temporal downsampling: assign a label to each stay using the partition returned by Infomap, and map labels back onto the input sequence. Non-stationary points are assigned label \textit{-1}, stationary ones are assigned a positive integer corresponding to a \emph{destination}.

    \item \texttt{SpatialInfostop}: \textbf{Cluster a collection of points using Infomap.} An instance of the class is initialised with the parameter $r2$ described in Section \ref{sec:algorithm}. The main method of the class is \texttt{fit\_predict}, which allows to cluster a collection of spatial points. The method entails points (b-e) described above.
    
\end{enumerate}

The package, furthermore, contains postprocessing utilities, that allow users to inspect various aspects of their solution, among these a visualization component that can visualize recovered destinations as convex hulls on top of a map, optionally with stay-points scattered or heatmapped (see Fig. \ref{fig:heatmap}).

Our implementation of Infostop is highly flexible. The default distance metric is the haversine distance, thus expecting the input location measurements to be \(lat, lon\) coordinates. However, the user can specify another distance function. As such, Infostop can be used as a general clustering algorithm in any space that allows distance measurement between points. The input data can be either in the form of one or multiple sequences of records, thus allowing usage for both single-user or multi-user datasets. Input sequences can include the timestamp associated to each record or not. Thus, it allows to use data sampled evenly or unevenly in time. 

\texttt{Infostop} is very performant. The temporal downsampling step, which entails iterating over the entire input sequence and computing running medians, is implemented in a C++ submodule for speed. Moreover, Infomap is itself a C++ library. Further speedups can be achieved by specifying a \texttt{min\_spacial\_resolution}, i.e. a lower bound on the spatial resolution. These speedups alone make multi-user stop-detection at the scale presented here feasible.

\section{Conclusion}
We have developed Infostop, an algorithm for stop-location detection that overcomes some of the limitations of the state-of-the-art stop-location detection algorithms, including that (i) frequently visited places located very close are likely to be merged into a unique location, due to inherent measurement noise, (ii) traces for multiple users can not be analysed simultaneously. We have shown that Infostop allows to identify shared stop-locations across a large population of individuals sharing several common locations. The outcome of the algorithm is more desirable than the output obtained using previous solutions, since locations maintain compact size. We showed that Infostop can be used to easily identify social contacts, or occasions when individuals spent simultaneously time in the same location. Furthermore, we have developed a fast implementation in C++ and Python, that is available open-source. The current implementation is highly flexible and faster, to our knowledge, than other open-source stop-location detection algorithms implementations. 
\bibliography{bibliography}

\begin{thebibliography}{18}%
\makeatletter
\providecommand \@ifxundefined [1]{%
 \@ifx{#1\undefined}
}%
\providecommand \@ifnum [1]{%
 \ifnum #1\expandafter \@firstoftwo
 \else \expandafter \@secondoftwo
 \fi
}%
\providecommand \@ifx [1]{%
 \ifx #1\expandafter \@firstoftwo
 \else \expandafter \@secondoftwo
 \fi
}%
\providecommand \natexlab [1]{#1}%
\providecommand \enquote  [1]{``#1''}%
\providecommand \bibnamefont  [1]{#1}%
\providecommand \bibfnamefont [1]{#1}%
\providecommand \citenamefont [1]{#1}%
\providecommand \href@noop [0]{\@secondoftwo}%
\providecommand \href [0]{\begingroup \@sanitize@url \@href}%
\providecommand \@href[1]{\@@startlink{#1}\@@href}%
\providecommand \@@href[1]{\endgroup#1\@@endlink}%
\providecommand \@sanitize@url [0]{\catcode `\\12\catcode `\$12\catcode
  `\&12\catcode `\#12\catcode `\^12\catcode `\_12\catcode `\%12\relax}%
\providecommand \@@startlink[1]{}%
\providecommand \@@endlink[0]{}%
\providecommand \url  [0]{\begingroup\@sanitize@url \@url }%
\providecommand \@url [1]{\endgroup\@href {#1}{\urlprefix }}%
\providecommand \urlprefix  [0]{URL }%
\providecommand \Eprint [0]{\href }%
\providecommand \doibase [0]{http://dx.doi.org/}%
\providecommand \selectlanguage [0]{\@gobble}%
\providecommand \bibinfo  [0]{\@secondoftwo}%
\providecommand \bibfield  [0]{\@secondoftwo}%
\providecommand \translation [1]{[#1]}%
\providecommand \BibitemOpen [0]{}%
\providecommand \bibitemStop [0]{}%
\providecommand \bibitemNoStop [0]{.\EOS\space}%
\providecommand \EOS [0]{\spacefactor3000\relax}%
\providecommand \BibitemShut  [1]{\csname bibitem#1\endcsname}%
\let\auto@bib@innerbib\@empty
\bibitem [{\citenamefont {Barbosa}\ \emph {et~al.}(2018)\citenamefont
  {Barbosa}, \citenamefont {Barthelemy}, \citenamefont {Ghoshal}, \citenamefont
  {James}, \citenamefont {Lenormand}, \citenamefont {Louail}, \citenamefont
  {Menezes}, \citenamefont {Ramasco}, \citenamefont {Simini},\ and\
  \citenamefont {Tomasini}}]{barbosa2018human}%
  \BibitemOpen
  \bibfield  {author} {\bibinfo {author} {\bibfnamefont {H.}~\bibnamefont
  {Barbosa}}, \bibinfo {author} {\bibfnamefont {M.}~\bibnamefont {Barthelemy}},
  \bibinfo {author} {\bibfnamefont {G.}~\bibnamefont {Ghoshal}}, \bibinfo
  {author} {\bibfnamefont {C.~R.}\ \bibnamefont {James}}, \bibinfo {author}
  {\bibfnamefont {M.}~\bibnamefont {Lenormand}}, \bibinfo {author}
  {\bibfnamefont {T.}~\bibnamefont {Louail}}, \bibinfo {author} {\bibfnamefont
  {R.}~\bibnamefont {Menezes}}, \bibinfo {author} {\bibfnamefont {J.~J.}\
  \bibnamefont {Ramasco}}, \bibinfo {author} {\bibfnamefont {F.}~\bibnamefont
  {Simini}}, \ and\ \bibinfo {author} {\bibfnamefont {M.}~\bibnamefont
  {Tomasini}},\ }\href@noop {} {\bibfield  {journal} {\bibinfo  {journal}
  {Physics Reports}\ }\textbf {\bibinfo {volume} {734}},\ \bibinfo {pages} {1}
  (\bibinfo {year} {2018})}\BibitemShut {NoStop}%
\bibitem [{\citenamefont {Blondel}\ \emph {et~al.}(2012)\citenamefont
  {Blondel}, \citenamefont {Esch}, \citenamefont {Chan}, \citenamefont
  {Cl{\'e}rot}, \citenamefont {Deville}, \citenamefont {Huens}, \citenamefont
  {Morlot}, \citenamefont {Smoreda},\ and\ \citenamefont
  {Ziemlicki}}]{blondel2012data}%
  \BibitemOpen
  \bibfield  {author} {\bibinfo {author} {\bibfnamefont {V.~D.}\ \bibnamefont
  {Blondel}}, \bibinfo {author} {\bibfnamefont {M.}~\bibnamefont {Esch}},
  \bibinfo {author} {\bibfnamefont {C.}~\bibnamefont {Chan}}, \bibinfo {author}
  {\bibfnamefont {F.}~\bibnamefont {Cl{\'e}rot}}, \bibinfo {author}
  {\bibfnamefont {P.}~\bibnamefont {Deville}}, \bibinfo {author} {\bibfnamefont
  {E.}~\bibnamefont {Huens}}, \bibinfo {author} {\bibfnamefont
  {F.}~\bibnamefont {Morlot}}, \bibinfo {author} {\bibfnamefont
  {Z.}~\bibnamefont {Smoreda}}, \ and\ \bibinfo {author} {\bibfnamefont
  {C.}~\bibnamefont {Ziemlicki}},\ }\href@noop {} {\bibfield  {journal}
  {\bibinfo  {journal} {arXiv preprint arXiv:1210.0137}\ } (\bibinfo {year}
  {2012})}\BibitemShut {NoStop}%
\bibitem [{\citenamefont {Stopczynski}\ \emph {et~al.}(2014)\citenamefont
  {Stopczynski}, \citenamefont {Sekara}, \citenamefont {Sapiezynski},
  \citenamefont {Cuttone}, \citenamefont {Madsen}, \citenamefont {Larsen},\
  and\ \citenamefont {Lehmann}}]{stopczynski2014measuring}%
  \BibitemOpen
  \bibfield  {author} {\bibinfo {author} {\bibfnamefont {A.}~\bibnamefont
  {Stopczynski}}, \bibinfo {author} {\bibfnamefont {V.}~\bibnamefont {Sekara}},
  \bibinfo {author} {\bibfnamefont {P.}~\bibnamefont {Sapiezynski}}, \bibinfo
  {author} {\bibfnamefont {A.}~\bibnamefont {Cuttone}}, \bibinfo {author}
  {\bibfnamefont {M.~M.}\ \bibnamefont {Madsen}}, \bibinfo {author}
  {\bibfnamefont {J.~E.}\ \bibnamefont {Larsen}}, \ and\ \bibinfo {author}
  {\bibfnamefont {S.}~\bibnamefont {Lehmann}},\ }\href@noop {} {\bibfield
  {journal} {\bibinfo  {journal} {PloS one}\ }\textbf {\bibinfo {volume} {9}},\
  \bibinfo {pages} {e95978} (\bibinfo {year} {2014})}\BibitemShut {NoStop}%
\bibitem [{\citenamefont {Gong}\ \emph {et~al.}(2015)\citenamefont {Gong},
  \citenamefont {Sato}, \citenamefont {Yamamoto}, \citenamefont {Miwa},\ and\
  \citenamefont {Morikawa}}]{Gong2015}%
  \BibitemOpen
  \bibfield  {author} {\bibinfo {author} {\bibfnamefont {L.}~\bibnamefont
  {Gong}}, \bibinfo {author} {\bibfnamefont {H.}~\bibnamefont {Sato}}, \bibinfo
  {author} {\bibfnamefont {T.}~\bibnamefont {Yamamoto}}, \bibinfo {author}
  {\bibfnamefont {T.}~\bibnamefont {Miwa}}, \ and\ \bibinfo {author}
  {\bibfnamefont {T.}~\bibnamefont {Morikawa}},\ }\href {\doibase
  10.1007/s40534-015-0079-x} {\bibfield  {journal} {\bibinfo  {journal}
  {Journal of Modern Transportation}\ }\textbf {\bibinfo {volume} {23}},\
  \bibinfo {pages} {202} (\bibinfo {year} {2015})}\BibitemShut {NoStop}%
\bibitem [{\citenamefont {Cao}\ \emph {et~al.}(2010)\citenamefont {Cao},
  \citenamefont {Cong},\ and\ \citenamefont {Jensen}}]{Cao2010}%
  \BibitemOpen
  \bibfield  {author} {\bibinfo {author} {\bibfnamefont {X.}~\bibnamefont
  {Cao}}, \bibinfo {author} {\bibfnamefont {G.}~\bibnamefont {Cong}}, \ and\
  \bibinfo {author} {\bibfnamefont {C.~S.}\ \bibnamefont {Jensen}},\ }\href
  {\doibase 10.14778/1920841.1920968} {\bibfield  {journal} {\bibinfo
  {journal} {Proc. VLDB Endow.}\ }\textbf {\bibinfo {volume} {3}},\ \bibinfo
  {pages} {1009} (\bibinfo {year} {2010})}\BibitemShut {NoStop}%
\bibitem [{\citenamefont {Montoliu}\ and\ \citenamefont
  {Gatica-Perez}(2010)}]{montoliu2010discovering}%
  \BibitemOpen
  \bibfield  {author} {\bibinfo {author} {\bibfnamefont {R.}~\bibnamefont
  {Montoliu}}\ and\ \bibinfo {author} {\bibfnamefont {D.}~\bibnamefont
  {Gatica-Perez}},\ }in\ \href@noop {} {\emph {\bibinfo {booktitle}
  {Proceedings of the 9th International Conference on Mobile and Ubiquitous
  Multimedia}}}\ (\bibinfo {organization} {ACM},\ \bibinfo {year} {2010})\
  p.~\bibinfo {pages} {12}\BibitemShut {NoStop}%
\bibitem [{\citenamefont {Do}\ and\ \citenamefont
  {Gatica-Perez}(2013)}]{do2013places}%
  \BibitemOpen
  \bibfield  {author} {\bibinfo {author} {\bibfnamefont {T.~M.~T.}\
  \bibnamefont {Do}}\ and\ \bibinfo {author} {\bibfnamefont {D.}~\bibnamefont
  {Gatica-Perez}},\ }\href@noop {} {\bibfield  {journal} {\bibinfo  {journal}
  {IEEE Transactions on Mobile Computing}\ }\textbf {\bibinfo {volume} {13}},\
  \bibinfo {pages} {638} (\bibinfo {year} {2013})}\BibitemShut {NoStop}%
\bibitem [{\citenamefont {Wan}\ and\ \citenamefont {Lin}(2013)}]{wan2013life}%
  \BibitemOpen
  \bibfield  {author} {\bibinfo {author} {\bibfnamefont {N.}~\bibnamefont
  {Wan}}\ and\ \bibinfo {author} {\bibfnamefont {G.}~\bibnamefont {Lin}},\
  }\href@noop {} {\bibfield  {journal} {\bibinfo  {journal} {Computers,
  Environment and Urban Systems}\ }\textbf {\bibinfo {volume} {39}},\ \bibinfo
  {pages} {63} (\bibinfo {year} {2013})}\BibitemShut {NoStop}%
\bibitem [{\citenamefont {Zhao}\ \emph {et~al.}(2015)\citenamefont {Zhao},
  \citenamefont {Ghorpade}, \citenamefont {Pereira}, \citenamefont {Zegras},\
  and\ \citenamefont {Ben-Akiva}}]{zhao2015stop}%
  \BibitemOpen
  \bibfield  {author} {\bibinfo {author} {\bibfnamefont {F.}~\bibnamefont
  {Zhao}}, \bibinfo {author} {\bibfnamefont {A.}~\bibnamefont {Ghorpade}},
  \bibinfo {author} {\bibfnamefont {F.~C.}\ \bibnamefont {Pereira}}, \bibinfo
  {author} {\bibfnamefont {C.}~\bibnamefont {Zegras}}, \ and\ \bibinfo {author}
  {\bibfnamefont {M.}~\bibnamefont {Ben-Akiva}},\ }\href@noop {} {\bibfield
  {journal} {\bibinfo  {journal} {Transportation research procedia}\ }\textbf
  {\bibinfo {volume} {11}},\ \bibinfo {pages} {218} (\bibinfo {year}
  {2015})}\BibitemShut {NoStop}%
\bibitem [{\citenamefont {Hariharan}\ and\ \citenamefont
  {Toyama}(2004)}]{hariharan2004project}%
  \BibitemOpen
  \bibfield  {author} {\bibinfo {author} {\bibfnamefont {R.}~\bibnamefont
  {Hariharan}}\ and\ \bibinfo {author} {\bibfnamefont {K.}~\bibnamefont
  {Toyama}},\ }in\ \href@noop {} {\emph {\bibinfo {booktitle} {International
  Conference on Geographic Information Science}}}\ (\bibinfo {organization}
  {Springer},\ \bibinfo {year} {2004})\ pp.\ \bibinfo {pages}
  {106--124}\BibitemShut {NoStop}%
\bibitem [{\citenamefont {Ester}\ \emph {et~al.}(1996)\citenamefont {Ester},
  \citenamefont {Kriegel}, \citenamefont {Sander}, \citenamefont {Xu} \emph
  {et~al.}}]{ester1996density}%
  \BibitemOpen
  \bibfield  {author} {\bibinfo {author} {\bibfnamefont {M.}~\bibnamefont
  {Ester}}, \bibinfo {author} {\bibfnamefont {H.-P.}\ \bibnamefont {Kriegel}},
  \bibinfo {author} {\bibfnamefont {J.}~\bibnamefont {Sander}}, \bibinfo
  {author} {\bibfnamefont {X.}~\bibnamefont {Xu}},  \emph {et~al.},\ }in\
  \href@noop {} {\emph {\bibinfo {booktitle} {Kdd}}},\ Vol.~\bibinfo {volume}
  {96}\ (\bibinfo {year} {1996})\ pp.\ \bibinfo {pages} {226--231}\BibitemShut
  {NoStop}%
\bibitem [{\citenamefont {Pappalardo}(2019)}]{pappalardo2019scikit}%
  \BibitemOpen
  \bibfield  {author} {\bibinfo {author} {\bibfnamefont {L.}~\bibnamefont
  {Pappalardo}},\ }\href@noop {} {\bibfield  {journal} {\bibinfo  {journal}
  {arXiv preprint arXiv:1907.07062}\ } (\bibinfo {year} {2019})}\BibitemShut
  {NoStop}%
\bibitem [{Note1()}]{Note1}%
  \BibitemOpen
  \bibinfo {note} {\protect \url
  {https://github.com/ulfaslak/infostop/blob/master/cpputils/main.cpp}}\BibitemShut
  {NoStop}%
\bibitem [{\citenamefont {Rosvall}\ \emph {et~al.}(2009)\citenamefont
  {Rosvall}, \citenamefont {Axelsson},\ and\ \citenamefont
  {Bergstrom}}]{rosvall2009map}%
  \BibitemOpen
  \bibfield  {author} {\bibinfo {author} {\bibfnamefont {M.}~\bibnamefont
  {Rosvall}}, \bibinfo {author} {\bibfnamefont {D.}~\bibnamefont {Axelsson}}, \
  and\ \bibinfo {author} {\bibfnamefont {C.~T.}\ \bibnamefont {Bergstrom}},\
  }\href@noop {} {\bibfield  {journal} {\bibinfo  {journal} {The European
  Physical Journal Special Topics}\ }\textbf {\bibinfo {volume} {178}},\
  \bibinfo {pages} {13} (\bibinfo {year} {2009})}\BibitemShut {NoStop}%
\bibitem [{Note2()}]{Note2}%
  \BibitemOpen
  \bibinfo {note} {\protect \url
  {https://github.com/ulfaslak/infostop/blob/master/infostop/utils.py}}\BibitemShut
  {NoStop}%
\bibitem [{\citenamefont {Steinley}(2004)}]{steinley2004properties}%
  \BibitemOpen
  \bibfield  {author} {\bibinfo {author} {\bibfnamefont {D.}~\bibnamefont
  {Steinley}},\ }\href@noop {} {\bibfield  {journal} {\bibinfo  {journal}
  {Psychological methods}\ }\textbf {\bibinfo {volume} {9}},\ \bibinfo {pages}
  {386} (\bibinfo {year} {2004})}\BibitemShut {NoStop}%
\bibitem [{Note3()}]{Note3}%
  \BibitemOpen
  \bibinfo {note} {\protect \url
  {https://mapequation.github.io/infomap/python/}}\BibitemShut {NoStop}%
\bibitem [{\citenamefont {Pedregosa}\ \emph {et~al.}(2011)\citenamefont
  {Pedregosa}, \citenamefont {Varoquaux}, \citenamefont {Gramfort},
  \citenamefont {Michel}, \citenamefont {Thirion}, \citenamefont {Grisel},
  \citenamefont {Blondel}, \citenamefont {Prettenhofer}, \citenamefont {Weiss},
  \citenamefont {Dubourg}, \citenamefont {Vanderplas}, \citenamefont {Passos},
  \citenamefont {Cournapeau}, \citenamefont {Brucher}, \citenamefont {Perrot},\
  and\ \citenamefont {Duchesnay}}]{scikit-learn}%
  \BibitemOpen
  \bibfield  {author} {\bibinfo {author} {\bibfnamefont {F.}~\bibnamefont
  {Pedregosa}}, \bibinfo {author} {\bibfnamefont {G.}~\bibnamefont
  {Varoquaux}}, \bibinfo {author} {\bibfnamefont {A.}~\bibnamefont {Gramfort}},
  \bibinfo {author} {\bibfnamefont {V.}~\bibnamefont {Michel}}, \bibinfo
  {author} {\bibfnamefont {B.}~\bibnamefont {Thirion}}, \bibinfo {author}
  {\bibfnamefont {O.}~\bibnamefont {Grisel}}, \bibinfo {author} {\bibfnamefont
  {M.}~\bibnamefont {Blondel}}, \bibinfo {author} {\bibfnamefont
  {P.}~\bibnamefont {Prettenhofer}}, \bibinfo {author} {\bibfnamefont
  {R.}~\bibnamefont {Weiss}}, \bibinfo {author} {\bibfnamefont
  {V.}~\bibnamefont {Dubourg}}, \bibinfo {author} {\bibfnamefont
  {J.}~\bibnamefont {Vanderplas}}, \bibinfo {author} {\bibfnamefont
  {A.}~\bibnamefont {Passos}}, \bibinfo {author} {\bibfnamefont
  {D.}~\bibnamefont {Cournapeau}}, \bibinfo {author} {\bibfnamefont
  {M.}~\bibnamefont {Brucher}}, \bibinfo {author} {\bibfnamefont
  {M.}~\bibnamefont {Perrot}}, \ and\ \bibinfo {author} {\bibfnamefont
  {E.}~\bibnamefont {Duchesnay}},\ }\href@noop {} {\bibfield  {journal}
  {\bibinfo  {journal} {Journal of Machine Learning Research}\ }\textbf
  {\bibinfo {volume} {12}},\ \bibinfo {pages} {2825} (\bibinfo {year}
  {2011})}\BibitemShut {NoStop}%
\end{thebibliography}%

\end{document}